# COMPARISON OF COVID19 PREDICTION PERFORMANCES OF NORMALIZATION METHODS ON COUGH ACOUSTICS SOUNDS

## Yunus Emre Erdoğan[1,2], Ali Narin[1]

[1]Electrical and Electronics Engineering Department, Zonguldak Bülent Ecevit University, 67100, Zonguldak, Turkey
[2]Electronic Automation Department, Eregli Iron and Steel Co., Zonguldak, Turkey

**Abstract**. *The disease called the new coronavirus (COVID19) is a new viral respiratory disease that first appeared on January 13, 2020 in Wuhan, China. Some of the symptoms of this disease are fever, cough, shortness of breath and difficulty in breathing. In more serious cases, death may occur as a result of infection. COVID19 emerged as a pandemic that affected the whole world in a little while. The most important issue in the fight against the epidemic is the early diagnosis and follow-up of COVID19 (+) patients. Therefore, in addition to the RT-PCR test, medical imaging methods are also used when identifying COVID 19 (+) patients. In this study, an alternative approach was proposed using cough data, one of the most prominent symptoms of COVID19 (+) patients. The performances of z-normalization and min-max normalization methods were investigated on these data. All features were obtained using discrete wavelet transform method. Support vector machines (SVM) was used as classifier algorithm. The highest performances of accuracy and F1-score were obtained as 100% and 100% using the min-max normalization, respectively. On the other hand, the highest accuracy and highest F1-score performances were obtained as 99.2 % and 99.0 % using the z-normalization, respectively. In light of the results, it is clear that cough acoustic data will contribute significantly to controlling COVID19 cases.*

1. INTRODUCTION

The novel coronavirus disease (COVID19) is a recent viral respiratory disease that was first identified on January 13, 2020 in Wuhan, China, with high fever and shortness of breath. It is known that the disease is transmitted by droplets and contact. It is defined as a pandemic because of the global epidemic situation it creates [1, 2]. The virus that induces the COVID19 pandemic is a serious acute respiratory syndrome coronavirus-2. (SARS-CoV-2) [3]. Symptoms of the new coronavirus infection can contain fever, cough, shortness of breath and difficulty breathing.

In more serious circumstances, the infection can induce pneumonia, acute respiratory insufficiency, kidney insufficiency and even decease [4]. It is also known that symptoms such as low lumbar pain, exhaustion, runs, queasiness, cephalagra and giddiness are seen [5].The identification of the virus can be made either directly to detect the virus, or by showing the specific antibodies that the host organism creates against the virus. In this direction, there are two test categories used in laboratories. These are the PCR test, which detects the virus itself, and the antibody test, which detects the host's response to the virus. In addition, lung tomography and some blood tests are utilized additionally in the identification of the illness [6]. These PCR tests have been annoying, time absorbing and those outcomes are lagged [7]. For this reason, approaches on the grounds that the analysis of cough acoustics sounds are utilized in together with this test [8]. Study on cough acoustics sound analysis were added to the literature.

Approaches on cough sounds for identifying COVID19 sicks are as indicated below: Erdoğan and Narin used cough acoustic data in their study. They used z-normalization technique for preprocessing. In the study, they used IMF and DWT based feature extraction via traditional learning techniques and they utilized SVM as classification method [9]. They also added deep features to this study. Here they used deep features on the RESNET50 deep learning model. In other research, Imran and colleagues tried to determine COVID19 sicks by examining cough sounds in a study named AI4COVID19. They engaged mean normalization technique in the research. They got images, which are Mel-spectrogram, for the Convolutional Neural Network pattern. For the traditional machine learning approach, they utilized Mel-frequency cepstral coefficients and Principal component analysis basis on feature extraction and Support Vector Machines classification method [10].

In this research, time domain and nonlinear features were taken via utilizing the 5-Layer discrete wavelet transform process utilizing cough sounds from people with COVID19(+) and COVID19(-) as shown in Figure 1. Z-normalization and min-max normalization were used as preprocessing method. In this study, which was carried out with the traditional machine learning approach, the support vector machine algorithm was used. In the second part of the study, the data set used, the details of the methods and techniques are given, the results obtained in the third part are given, while the findings obtained from the studies in the literature are discussed in the fourth part.

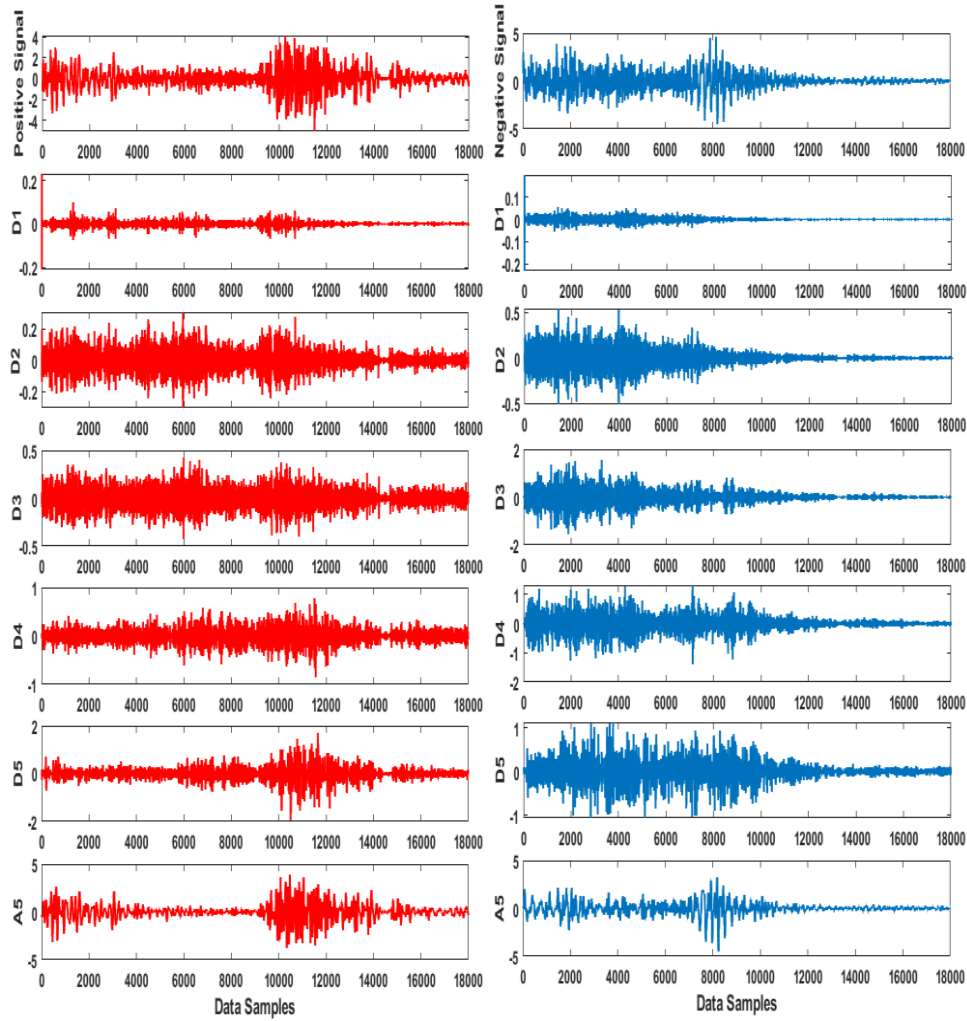

**Figure 1.** COVID19 (+) and COVID19 (-) cough acoustics data and 5-Layer Discrete Wavelet Transform coefficients

## 2.   METHODS

### 2.1. Dataset

Cough sounds of people who are COVID19 (+) and are COVID19 (-) have been taken from https://virufy.org/. The data was collected with a mobile based application builded up via Stanford University. The cough sounds pertain to a sum of 16 people and 6 of these people are females and 10 of them are males. These group of people has 42 age as average.

4    Y.E. ERDOĞAN, A. NARINAll data have been stated as positive and negative pursuant to the outcomes collected from PCR test. As a conclusion of these tests, COVID 19 (+) was 7 of these group of people and 9 of them were COVID 19 (-).These results were tagged. Most people tagged as COVID19 (+) have signs like difficulty in breathing, sore throat, and cough. Some of them suffer from congestive heart failure, asthma, diabetes, and in some others have a loss of taste and smell, fever, and chills. Only one of the people labeled COVID19 (-) have chronic diabetes and a few have complaints of difficulty of breathing, throat ache, body aches and worsening cough. The data on the platform from which it was obtained was segmented and made ready for the use of researchers. From these data, 9 COVID19 (-) labeled cough acoustic data were divided into 73 parts, 7 COVID19 (+) labeled cough acoustic data were divided into 48 parts. The sampling frequency of each piece of data is 48 kHz, and the coughing time has been determined as 1640 milliseconds (ms). In addition, all the data were standardized with the z-score method before the study carried out.

### 2.2. Z-Normalization

Z-score converts mean of variable to 0 and its standard deviation to 1. To do this, simply subtract the mean and divide by the standard deviation. The z-score normalization can be calculated as [11]:

$$z = \frac{x - m}{sd} \tag{1}$$

Where x indicates any point in the dataset, m indicates mean of the dataset, and sd indicates the standard deviation of the dataset.

### 2.3. Min-Max Normalization

In the min-max normalization process, the features are normalized between 0 and 1 according to the equation below [11].

$$v' = \frac{v - min_X}{max_X - min_X} \tag{2}$$

Here, $min_x$ is the minimum value of the x feature and $max_x$ is the maximum value of the x feature. The original and normalized values of the x feature are denoted by v ve v', respectively. It can be seen from the above equation that the minimum and maximum attribute values are equal to 0 and 1, respectively.

### 2.4 Discrete Wavelet Transform

Process of the continuous wavelet transform is the sum of the signal which is multiplied by its shifted and scaled types in the time plane of the main wavelet during time. As a conclusion of these operations, wavelet coefficients based upon the position and the scale are found. Provided that scaling and translation are preferred as powers of 2 analyzes are more effective than continuous wavelet transform and gives as accurate results as it. This kind of analysis is called the discrete wavelet transform [12].

### 2.5 Support Vector Machines

SVM is a classification method on the grounds that statistical theory. The mathematical algorithms of the SVM were initially designed for the classification problem of two-class linear data, then generalized for the classification of multi-class and non-linear data. The working principle of SVM is based on estimating the most appropriate decision function that can distinguish two classes from each other, in other words, defining the hyper-plane that can best distinguish two classes from each other [13].

### 2.6 Performance Metrics

In this study, the results were evaluated using five different performance measures [14, 15]. These:

$$Accuracy(ACC) = \frac{TP + TN}{TP + FN + FP + TN} \quad (3)$$

$$Recall(REC) = \frac{TP}{TP + FN} \quad (4)$$

$$Specifity(SPE) = \frac{TN}{TN + FP} \quad (5)$$

$$Precision(PRE) = \frac{TP}{TP + FP} \quad (6)$$

$$F1 - score(F1) = \frac{2 * PRE * REC}{PRE + REC} \quad (7)$$

TP=True Positive, FP=False Positive, TN=True Negative, FN=False Negative.

TP shows the people ,who are COVID19(+), number and identified as COVID19(+) by the classifier, FN is the number of people who are wrongly stated as COVID19(-), TN the people number, who are actually COVID19(-),and the classifier were stated them as COVID19(-), and FP shows the people number ,who are mistakenly identified as COVID19(+) [16].

### 3. EXPERIMENTAL RESULTS

In this research work, the preprocessing of the cough sound signals taken, the features taken with traditional machine learning methods, extraction operation of the features obtained and taken of the performance values have been experimented via MATLAB 2020 computer programme. Both normalization methods utilized in this research, traditional machine learning processes have been engaged. Then, 54 DWT based measures were found from the signals taken as a conclusion of min-max normalization and z-score. The confusion matrix of z-normalization, the highest performance result from the results obtained with the features extracted by traditional approaches is given in Figure 2. It is clearly seen that while all the ones in the COVID19 (+) class were detected correctly, only 1 of the ones in the COVID19 (-) class were detected incorrectly. For min-max normalization, all of the COVID19 (+) and COVID19 (-) classes have been correctly identified.

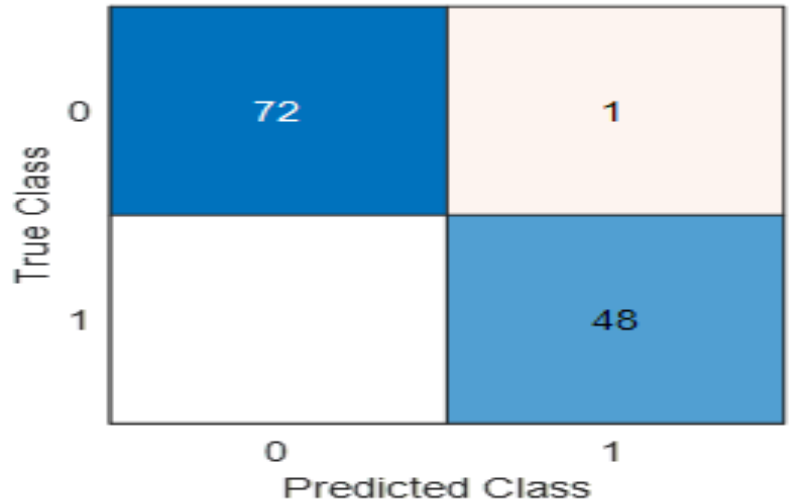

**Figure 2.** Z-score confusion matrice with highest performance

## 4. DISCUSSION

No doubt one of the most researched topics in the last few years is the coronavirus epidemic, which affected all over the world. The most vital point in overcoming this epidemic is the process of making the correct diagnosis. Therefore, there are many machine learning-based studies in the literature [17]. The detection of COVID19 (+) through cough acoustic signals, which suggests a different approach, has recently taken its place among the current alternative methods. Performance metrics comparisons of those using normalization methods among the studies conducted in this area are given in Table 1. Traditional machine learning approaches and feature extraction processes were generally used in these studies. In some studies, in addition to traditional machine learning approaches, deep learning approaches have been used to detect COVID19 (+) people based on cough sounds. In this study, the performances of two different normalization methods were analyzed over the features obtained with traditional machine learning approaches. With these two alternative approaches, which work with a very high success rate, a decision support mechanism has been recommended to experts for the detection of COVID19 (+) people.

In addition to the detection of COVID19 (+) with imaging methods, it is of great importance to detect these people with cough-based acoustic sound analysis. With this method, the detection of COVID19 (+) can be easily achieved via a smartphone or computer application. With this application, the pandemic can be overcome more easily. From this point of view, it is of great importance that even a single person can be protected from the epidemic during the pandemic. We think that such systems showing high performance will be of importance during pandemic period. One of the most critical restrictions of this research is the limited number of data. By enhancing the number of data near future, it is thought that the system will be successful on high data numbers. In future studies, it is planned to increase the number of nonlinear measurements.

**Table 1.** COVID19 detection with normalization techniques using acoustics signals in the literature.

| Authors | Methods and Classifiers | Number of Data | Performance (%) |
| --- | --- | --- | --- |
| Imran et al. (2020) [10] | Mean Score + Mel-frequency cepstral coefficients and Principal component analysis / Support Vector Machines | 543 | Rec=96.0 Spe=95.2 Acc=95.6 F1=95.6 |
| Erdogan and Narin(2021) [9] | Z-Score + Intrinsic Mode Functions and Discrete Wavelet Transform features + Support Vector Machines | 1187 | Rec=99.5 Spe=97.4 Acc=98.4 F1=98.6 |
| Erdogan and Narin(2021) [9] | Z-Score + ResNet50 basis deep features + Support Vector Machines | 1187 | Rec=98.5 Spe=97.3 Acc=97.8 F1=98.0 |
| **This study** | **DWT based features + Min-max Score/Support Vector Machines** | **121** | **Rec=100 Spe=100 Acc=100 F1=100** |
| **This study** | **DWT basis features + Z Score / Support Vector Machines** | **121** | **Rec=100 Spe=98.6 Acc=99.2 F1=99.0** |